%% file: main.tex
\newcommand{\etal}{\textit{et al.}}

\documentclass[conference,compsoc]{IEEEtran}
\IEEEoverridecommandlockouts

\ifCLASSOPTIONcompsoc
  \usepackage[nocompress]{cite}
\else
  \usepackage{cite}
\fi
\ifCLASSINFOpdf
\else
\fi
\usepackage{amsmath}
\usepackage{xurl}
\usepackage{hyperref}
\usepackage{booktabs}
\usepackage{multirow}
\usepackage{algorithmic}

\usepackage{array}

 \usepackage[caption=false,font=footnotesize,labelfont=sf,textfont=sf]{subfig}
\usepackage{url}

\usepackage{graphicx}

\usepackage{tcolorbox}
\usepackage{xcolor}
\usepackage{amsmath}
\usepackage{graphicx}

\begin{document}
%
\title{Mental Damage: Caption Poisoning Attacks on Retrieval-Augmented Text-to-Music Generation}




\author{
\IEEEauthorblockN{
Yizhu Wen\IEEEauthorrefmark{1}\IEEEauthorrefmark{4},
Shuhao Zhang\IEEEauthorrefmark{2}\IEEEauthorrefmark{4},
Nan Zhang\IEEEauthorrefmark{3},
Long Cheng\IEEEauthorrefmark{2}, 
Hanqing Guo\IEEEauthorrefmark{1}
}

\IEEEauthorblockA{\IEEEauthorrefmark{1}
University of Hawai\textquotesingle i at M\=anoa, 
\{yizhuw, guohanqi\}@hawaii.edu}

\IEEEauthorblockA{\IEEEauthorrefmark{2}
Clemson University,
\{shuhaoz, lcheng2\}@clemson.edu
}

\IEEEauthorblockA{\IEEEauthorrefmark{3}
Michigan State University, 
\{zhangn24\}@msu.edu}

\thanks{\IEEEauthorrefmark{4} Yizhu Wen and Shuhao Zhang contributed equally to this work.}
}


\maketitle
\renewcommand{\thefootnote}{\fnsymbol{footnote}} 

\input{section/0_abstract}
\input{section/1_intro}

\input{section/2_background}

\input{section/3_threat_model}

\input{section/4_design}
\input{section/5_evaluation}

\input{section/6_discussion}
\input{section/8_conclusion}

\bibliographystyle{IEEEtran}
%
\bibliography{bib}
\input{section/9_appendix}
\end{document}

%% file: section/0_abstract.tex

\begin{abstract}
Retrieval-augmented text-to-music (TTM) systems augment underspecified user prompts using captions retrieved from a music caption dataset. This design introduces an integrity dependency on the music knowledge database. We show that an attacker can poison the database by injecting a small number of crafted music captions, causing the system to retrieve malicious captions that bias prompt augmentation and steer generation away from the user’s intended function, without modifying the user prompt, retriever, or generator. To achieve the music caption poisoning attack, we propose a dual-layer caption poisoning strategy that preserves high-level retrieval anchors while injecting low-level acoustic descriptors to steer prompt augmentation and downstream music generation toward an attacker-chosen target intent. In a MusicCaps knowledge database, CLAP retriever, and MusicGen pipeline, poisoned generations move substantially closer to the attacker's target, while remaining comparably aligned with the original user query. These results expose a practical integrity risk for retrieval-augmented creative AI systems. Our demo can be found at: \url{https://yizhu-wen.github.io/Mental-Damage/}
\end{abstract}

%% file: section/1_intro.tex
\section{Introduction}
Text-to-music (TTM) generation systems, such as MusicLM \cite{agostinelli2023musiclm}, MusicGen \cite{huang2023musicgen}, and AudioLDM \cite{liu2023audioldm}, have made music creation increasingly accessible by allowing users to specify desired audio content in natural language. In practice, however, models trained on low-level domain-specific semantics \cite{agostinelli2023musiclm} struggle with underspecified real-world queries that people usually ask \cite{doh2024music}. For example, a user without professional training in the music domain may write a high-level prompt such as \emph{``Give me a calm song to support my self-study''} or \emph{``I need relaxing background music for reading''} \cite{knees2013survey}. These prompts clearly express user intent (e.g., calmness, focus, study support), but they provide insufficient low-level acoustic guidance (e.g., tempo, instrumentation, timbre, texture, dynamics) for high-quality music generation. Recent work~\cite{gonzales2024retrieval, ding2025ragrewrite} addresses this gap with retrieval-augmented prompt augmentation and rewriting, where the system retrieves captions or examples from a music-text corpus and uses them to enhance the user prompt into a more detailed, expert-like description before generation. For instance, the system may rewrite a high-level prompt into a low-level description such as \emph{``slow tempo piano-led ambient track, soft dynamics, sparse arrangement, warm pad texture, low rhythmic complexity, gentle attack, no abrupt transients''}. This design can improve usability and output quality, but it also introduces a new security dependency, the integrity of the retrieval corpus and, in particular, the textual labels/captions attached to the audio.

Prior work on retrieval-augmented generation (RAG) security has shown that RAG systems can be attacked by poisoning the retrieval knowledge base, causing downstream models to produce attacker-influenced outputs \cite{zou2025poisonedrag,lewis2020rag}. Existing RAG poisoning studies primarily focus on text-centric LLM settings, where the attack objective is to manipulate factual answers or textual completions. It remains unclear whether similar attacks transfer to multimodal generation pipelines, especially TTM systems, where retrieved captions directly guide generation.

\begin{figure}[t]
\centering
\includegraphics[width=\linewidth]{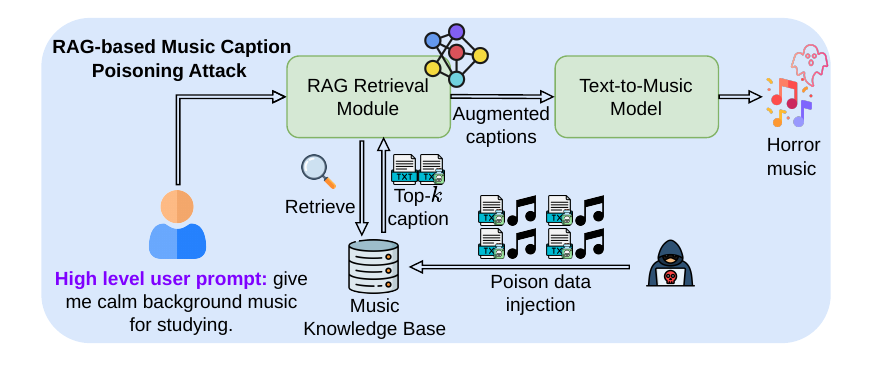}
    \vspace{-20pt}
\caption{RAG-based caption augmentation for text-to-music generation: enrich a high-level user prompt by retrieving detailed, low-level music captions from a music–text knowledge base, then pass the augmented prompt to a TTM model.}
    \vspace{-15pt}
\label{fig:rag_prompt_rewriting}
\end{figure}

In this paper, we study a new attack surface in retrieval-augmented text-to-music (TTM) generation, namely, \emph{caption poisoning of the music RAG knowledge database}. Figure~\ref{fig:rag_prompt_rewriting} shows the victim pipeline and the poisoning entry point. A user provides a \emph{high-level prompt} (for example, “calm song for study”). The system embeds this query and retrieves the most relevant captions from a music-text knowledge base. The retrieved captions are then fed to a \emph{prompt augmentation module}, either via direct concatenation or an LLM-based rewriter, which produces a \emph{low-level prompt} describing generation-oriented acoustic attributes such as tempo, instrumentation, timbre, and texture. The TTM generator (e.g., MusicLM) then synthesizes audio conditioned on this augmented prompt.


\textbf{Attack Goal and Impact:} The attacker's goal is to induce a RAG-based TTM system to generate music that deviates from the user's intended function, even when the user provides a benign query. Instead of attacking the generator model directly, the attacker targets the RAG knowledge database so that the system receives manipulated caption guidance during prompt augmentation. If successful, the attack causes a mismatch between user intent and generated output (e.g., producing music with unintended affective or stylistic characteristics), while leaving the inference-time user prompt, retriever architecture, and generator model parameters unchanged. This makes the attack difficult to notice from the user's perspective and exposes a practical integrity risk in retrieval-augmented creative generation pipelines.


\textbf{Challenges:} However, launching such an attack is non-trivial because a successful poisoned caption must satisfy multiple competing requirements. \textbf{(1) Retrieval Plausibility:} the poisoned caption must remain semantically relevant enough to be retrieved for the target benign query (e.g., if a user asks for \emph{``calm music for studying''}, the poisoned caption must still look related to calm/study/background music). \textbf{(2) Conflict Control:} naive poisoning may introduce target semantics that are too strongly conflicting, which can reduce retrieval plausibility or create unstable conditioning for the generator. For example, directly appending a target label such as \emph{``rock''} to a calm-study caption may conflict with the source intent in both high-level function and low-level acoustic traits (e.g., tempo, dynamics, and rhythmic density). \textbf{(3) Steering Strength:} the poisoned caption must also carry sufficient malicious guidance to influence prompt rewriting and downstream generation (e.g., shifting the augmented prompt toward eerie or spooky acoustic characteristics).

\textbf{Our Idea:} Prior work on automated audio caption analysis separates caption content into high-level intent elements and low-level acoustic descriptors \cite{xu2023beyond}. Building on this distinction, we address the retrieval steering tension by decoupling \emph{high-level retrieval compatibility} from \emph{low-level generation steering} via a three-part design. \textbf{(1) Anchor Preservation} improves \emph{retrieval plausibility} by retaining high-level semantic cues that keep poisoned captions retrievable under benign queries, for example, \emph{calm study music''} or \emph{relaxing background music''}. \textbf{(2) High-level Functional Opposite Target Generation} improves \emph{conflict control} by selecting target concepts that invert the user’s intended function (e.g., comfort/focus vs.\ unease/tension), while remaining compatible in the low-level descriptor space, which reduces instability relative to strongly contradictory targets. \textbf{(3) Low-level Semantic Similar Target Generation} improves \emph{steering strength} by injecting attacker-controlled low-level acoustic descriptors that shape prompt augmentation and downstream synthesis, such as \emph{slow pulsing drone,'' distant echoing chimes,''} or \emph{``hollow reverberant space''}. Together, these components produce poisoned captions that remain plausible at retrieval time while redirecting generation away from the user’s intended function. We summarize our Contributions below:
\begin{itemize}
    \item We identify a new integrity attack against RAG-based TTM generation, where poisoning caption metadata in the RAG corpus can redirect the high-level-to-low-level prompt translation process and cause generated music to deviate from the user's intended function.
    \item We design a dual-level caption poisoning attack that addresses the key attack challenges by combining \emph{Anchor Preservation}, \emph{Function-Opposed Targeting}, and \emph{Descriptor-Level Payload Injection} to jointly improve retrieval plausibility and malicious generation steering.
    \item We demonstrate the effectiveness of the attack on a RAG-based TTM pipeline using CLAP retrieval and MusicGen music generation, showing that poisoned generated music can roughly double similarity to an attacker-chosen target intent while preserving alignment with the original user query.
\end{itemize}

%% file: section/2_background.tex
\section{Background}
\subsection{What is Text-to-Music (TTM)?}
TTM refers to a class of generative AI systems that create music from natural-language prompts. Representative examples include MusicLM \cite{agostinelli2023musiclm}, MusicGen \cite{huang2023musicgen}, and audio-generation models closely related to TTM, such as AudioLDM \cite{liu2023audioldm}. In a TTM system, a user describes the desired music in text, such as mood (e.g., calm, tense), genre (e.g., classical, jazz), instrumentation (e.g., piano, guitar, strings), tempo (e.g., slow, fast), or scene context (e.g., background music for studying), and the model generates a corresponding audio clip.
Intuitively, TTM acts as a translation system between two different modalities: human language and music audio. The text prompt provides a semantic description of what the user wants, and the model converts that description into musical signals. Systems such as MusicLM \cite{agostinelli2023musiclm} and MusicGen \cite{huang2023musicgen} illustrate this text-conditioned generation setting, while audio-text alignment models such as CLAP \cite{elizalde2023clap} support stronger semantic grounding between text and audio representations. To do this well, a TTM model must not only understand high-level intent in the prompt (e.g., relaxing vs.\ energetic), but also map that intent into lower-level musical attributes such as rhythm, timbre, texture, dynamics, and instrumentation. In this sense, TTM can be viewed as a multimodal conditional generation task, where text serves as the control signal and music is the generated output \cite{huang2023musicgen,liu2023audioldm}.
TTM is challenging because user prompts are often high-level and underspecified. For example, a prompt like ``calm music for studying'' conveys intent, but omits generation-relevant attributes such as instrumentation, arrangement density, tempo, and spatial texture. To reduce this gap, recent systems leverage stronger text-audio alignment, richer caption supervision (for example, MusicCaps in the MusicLM pipeline \cite{agostinelli2023musiclm,musiccaps_hf}), and prompt refinement mechanisms such as retrieval-augmented prompt rewriting \cite{ding2025ragrewrite} to improve controllability and generation quality.

\subsection{RAG for TTM}
RAG, originally popularized in LLM-based text generation, has been increasingly adopted in multimodal generation settings, including text-to-image and text-image generation~\cite{blattmann2022retrieval,sheynin2022knn,pmlr_yasunaga}. With recent advances in large-scale generative audio models, similar retrieval-augmented mechanisms have also emerged in text-to-audio (TTA) and TTM generation~\cite{ghosh2024recap,changin2024enhancing}. These systems use external audio-text memory (e.g., captioned music or audio examples) to improve semantic grounding, controllability, and generation quality, especially for abstract prompts, long-tailed categories, and zero-shot settings.

In the TTA setting, prior work has incorporated retrieval into diffusion and flow-matching pipelines. Yuan~\etal~\cite{ghosh2024recap} integrate CLAP-based retrieval into latent diffusion models to improve the generation of rare or unseen sounds. Yang~\etal~\cite{yang2024audiobox} condition flow-matching generation on retrieved audio exemplars, improving zero-shot and few-shot performance without requiring labeled retrieval data. Zhao~\etal~\cite{zhao2025feedback} further introduce a feedback-driven refinement framework using large audio language models, where missing or imperfect sound events are iteratively identified and corrected through retrieval and cross-attention.

In the music domain, retrieval is also applied at the prompt level to improve controllability and expressiveness in TTM systems. Gonzales-Rudzicz~\etal~\cite{gonzales2024retrieval} augment input prompts with semantically similar musical aspects retrieved from MusicCaps~\cite{musiccaps_hf}. Ding~\etal~\cite{ding2025ragrewrite} retrieve relevant descriptions and use a language model to rewrite novice prompts into more detailed, expert-style inputs, improving musicality and production quality without modifying the generator. Collectively, these approaches reveal a common design pattern in RAG-enabled TTM systems: retrieved captions function as an intermediate semantic control signal that bridges high-level user intent and low-level acoustic realization.


\noindent\textbf{Why TTM Performs Well on Low-Level Input?}
TTM models often perform well on low-level acoustic descriptions (e.g., tempo, texture, reverberation, timbre, and rhythmic density) because such descriptors are frequently observed in music caption datasets, whereas high-level semantic intents are more diverse and relatively sparse \cite{sun2021research}. This repeated supervision makes low-level descriptors stable and effective conditioning signals during training \cite{xu2023beyond}. By contrast, high-level prompts are often underspecified and can be realized in many musically different ways. For example, ``music for studying'' may correspond to different genres, instruments, and arrangement styles. As a result, low-level descriptors often provide stronger control over generation.

\subsection{Existing RAG Attacks}
The security risks of RAG have received increasing attention as RAG has been widely integrated into large language model systems. Prior work has shown that RAG can be attacked in both text-based and image-based settings~\cite{Guu_2020,lewis2020rag,pmlr_yasunaga, Meng2025}. In text-based RAG, attacks often manipulate the external knowledge source (e.g., vector database or retrieved documents) through poisoning or injection, causing the system to retrieve malicious content that steers downstream generation~\cite{zou2025poisonedrag}. In image-based or multimodal RAG settings, attackers can exploit cross-modal alignment and retrieval mechanisms to bias what the system retrieves and, consequently, what it generates~\cite{zhangpoisonedeye,liu2025poisoned}.
These results establish an important security lesson: in RAG systems, the attack surface is not limited to the generative model itself, but also includes the external retrieval corpus and the embedding space that links user queries to retrieved content. However, existing studies have largely focused on text and image domains. In contrast, the security of \emph{audio-based} and \emph{text-to-music} RAG pipelines remains underexplored, despite the increasing use of audio-text retrieval, caption-guided prompt rewriting, and cross-modal audio-text embeddings in modern TTM systems.

%% file: section/3_threat_model.tex
\section{Threat Model}

\begin{table*}[ht]
\centering
\vspace{-6pt}
\caption{Glossary and notation for the music caption poisoning attacks}
\vspace{-6pt}
\label{tab:prompt_terms}
\renewcommand{\arraystretch}{1.25}
\setlength{\tabcolsep}{10pt}
\begin{tabular}{p{3.6cm} p{2.2cm} p{9.7cm}}
\toprule
\textbf{Concept} & \textbf{Notation} & \textbf{Meaning} \\
\midrule

Benign target query & $q$ &
Natural-language request expressing intent only, without technical audio descriptions. \\

Target category & $t$ &
Functionally opposite category that shares similar low-level acoustics with the benign user intent. \\

High-level user intent anchor & \(c_\text{anchor}\) &
High-level safe listening goal representing the original user intention. \\

Low-level semantic payload & \(c_\text{payload}\) &
Low-level musical attributes common to both intents explain acoustic similarity. \\

Target music captions & $\{c_p\}_{i=1}^N$ &
A set of \(N\) captions that describe low-level acoustic attributes and are intended to align with the attacker’s target category. \\
\bottomrule
\end{tabular}
\vspace{-10pt}
\end{table*}

\noindent\textbf{Victim setting.} We consider a victim who constructs a music RAG system by collecting music-caption pairs from public data sources and ingesting them into a local knowledge base. At inference time, a user issues a textual query to request the TTM model to generate a music clip. The system performs text-to-text retrieval by comparing the query embedding against caption embeddings stored in the knowledge base and returns the most similar captions. The TTM model then generates a music clip conditioned on both the user query and the retrieved captions. 


\noindent\textbf{Attacker’s capability.} We assume that the attacker can only inject malicious music-text (music-caption) pairs into the public data sources. These poisoned entries are subsequently collected and incorporated into the victim’s knowledge database during data ingestion. This threat model is consistent with prior work on poisoned RAG systems, such as PoisonedRAG \cite{zou2025poisonedrag}, and PoisonedEye \cite{zhangpoisonedeye}.

\noindent\textbf{Attacker’s goal.} The attacker aims to cause a RAG-based TTM system to generate music that deviates from the benign user’s intended function and instead follows an attacker-chosen target intent \(t\). Given a benign query \(q\) (e.g., music for studying or therapy), the attacker selects a high-level, functionally opposite target category (e.g., horror or haunted music) but has acoustic commonalities (e.g., slow decay, deep resonant) and seeks to induce the TTM model to generate music belonging to the attacker-chose target cateegory.

For example, a user may request “calm background music for studying,” while the attacker aims to trigger the generation of horror-style music reminiscent of a haunted movie. The attacker achieves this by corrupting the RAG's caption guidance music generation process in the knowledge base, without modifying the user’s query. Such attacks can disrupt creative workflows, replace intended audio with harmful or inappropriate content, and introduce reputational and safety risks in RAG-based music generation systems.

\noindent\textbf{Attacker capability.} We consider a black-box attacker with no access to the deployed knowledge base, retriever, or model internals, including the TTM model parameters. For each benign query \(q\), the attacker can inject up to \(N\) malicious music-caption pairs into public datasets that the victim later collects. These injected pairs can be sourced from publicly available repositories, such as MusicCaps \cite{agostinelli2023musiclm}.

%% file: section/4_design.tex
\section{Design}
Music caption datasets typically follow a long-tailed distribution: low-level acoustic descriptors (e.g., tempo, texture, reverberation) are abundant, whereas high-level semantic categories are relatively sparse~\cite{sun2021research, xu2023beyond}. As a result, TTM models trained on such data tend to rely heavily on low-level cues when conditioned on text prompts. In music RAG systems, this tendency is further amplified: semantically similar descriptions are retrieved from external corpora and incorporated into user prompts. As caption retrieval operates in a shared embedding space, high-level category features can be approximated through low-level acoustic attributes. Building on this observation, we propose a Caption Poisoning attack targeting the prompt retrieval stage of the RAG-based TTM system. The attacker selects a benign source category (e.g.,\emph{`` music used in spa or study environments''}) and a conflicting target category (e.g., \emph{``horror film background music''}). Although these categories differ semantically at a high level, they may share low-level characteristics, such as sustained drones, slow tempo, minimal motion, and heavy reverberation. By describing the target style using low-level descriptors that resemble the source category in the retrieval embedding space, the attacker can craft adversarial captions and inject them into the RAG knowledge database. When the system retrieves context for a benign user query, these poisoned captions can enter the top-k TTM context and steer generation toward the attacker-chosen target category.

The core of our attack is poisoned caption crafting. A successful poisoned caption must remain retrievable for benign queries, while also carrying malicious guidance that steers downstream music generation. This creates an inherent tradeoff: adding malicious content can reduce the caption’s semantic similarity to the benign queries, thus reducing the likelihood of being retrieved. We address this retrieval--steering tradeoff with a design that separates high-level retrieval compatibility from low-level generation steering. Our attack consists of three components: (1) Anchor Preservation, (2) Function-Opposed Targeting, and (3) Descriptor-Level Payload Injection.

\noindent\textbf{Problem Formulation:}
Given a benign target query $q$, our goal is to construct a poisoned caption $c_p$ that is likely to be retrieved for $q$ while steering downstream generation toward an attacker-chosen target category $t$. We formulate the poisoned caption as:
\begin{equation}
c_p = c_{\text{anchor}}(q) \Vert c_{\text{payload}}(t),
\label{eq:cp_basic}
\end{equation}
where $c_{\text{anchor}}(q)$ preserves query-relevant semantics for retrieval plausibility, $c_{\text{payload}}(t)$ provides malicious generation guidance, and $\Vert$ denotes text concatenation. To make the retrieval--steering tradeoff explicit, we view poisoned caption crafting as optimizing two objectives:
\begin{equation}
\max_{c_p} \;\; \mathcal{R}(c_p; q) + \lambda \,\mathcal{S}(c_p; t),
\label{eq:objective}
\end{equation}
where $\mathcal{R}(\cdot)$ measures retrieval plausibility given the benign query $q$, $\mathcal{S}(\cdot)$ measures steering strength toward target category $t$, and $\lambda$ controls the retrieval--steering tradeoff. Table~\ref{tab:prompt_terms} summarizes the concepts and notation used in our music caption poisoning formulation, clarifying how the benign query $q$, target category $t$, and poisoned caption components $(c_\text{anchor}, c_\text{payload}, \{c_p\}_{i=1}^N)$ are defined and referenced throughout the paper.

\begin{figure*}[ht]
\vspace{-15pt}
\centering
\includegraphics[width=\linewidth]{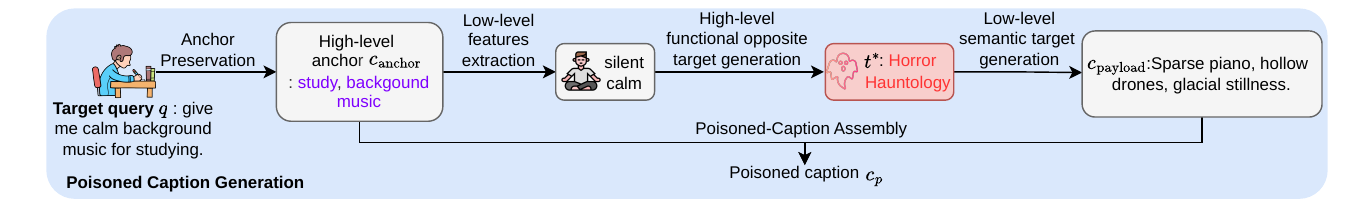}
    \vspace{-15pt}
\caption{Overview of the poisoned caption generation pipeline. Given a target query \(q\), the system first extracts a high-level function anchor \(c_{\text{anchor}}\) and then derives low-level musical attributes from the query. Conditioned on these attributes, it selects a functionally opposite high-level target \(t^\ast\). Finally, it generates a low-level semantic payload \(c_{\text{payload}}\) consistent with \(t^\ast\). The poisoned caption is formed by concatenating \(c_\text{anchor}\) and \(c_\text{payload}\).}
\label{fig:system}
    \vspace{-15pt}
\end{figure*}

\subsection{Anchor Preservation}
This component addresses \emph{retrieval plausibility}. The key idea is simple: the poisoned caption must still appear to match the benign user query; otherwise, it will not be retrieved. To achieve this, we keep a query-aligned \emph{anchor} \(c_{\text{anchor}}\) in the poisoned caption that preserves the source category at a high level. For example, if the benign query is about calm music for studying, the poisoned caption should still contain semantics such as \emph{``calm study music''} or \emph{``relaxing background music''}. These anchor cues help the caption remain relevant during retrieval. Intuitively, Anchor Preservation keeps the poisoned caption ``close enough'' to the benign query in high-level semantic space so that the retriever continues to treat it as a plausible match. This is the part of the caption that helps the attack get selected before any malicious steering can take effect.

\subsection{High-level Functional Opposite Target Generation}
This component addresses \emph{conflict control}. The goal is to choose a target category that changes the \emph{function} of the generated music (e.g., from comfort/focus to unease/tension), but does not introduce such a strong low-level mismatch that retrieval or generation becomes unstable.

The key intuition is that not all target styles are equally effective for poisoning. Some targets are simply too contradictory. For example, if the benign query is about \emph{calm study music}, a target category such as \emph{rock music} conflicts at both a high level (intended function) and a low level (e.g., tempo, rhythmic density, and dynamics), making the poisoned caption less retrievable and the poisoned generation less stable. In contrast, a target such as \emph{spooky} can still oppose the user's intended function while sharing low-level traits with calm music (e.g., slow tempo, sparse texture, and sustained sounds). We reconcile the tradeoff by selecting a target that is functionally opposed to the source category while avoiding excessive conflict:

\begin{equation}
t^\ast = \arg\max_{t \in \mathcal{T}} \Big( \mathrm{Oppose}_{\text{func}}(s,t) - \,\mathrm{Conflict}_{\text{desc}}(s,t) \Big),
\label{eq:target_select}
\end{equation}

where $s$ denotes the source category of the benign query \(q\), and $t^\ast$ is the attacker-chosen category. $\mathrm{Oppose}_{\text{func}}(s,t)$ scores how strongly a candidate target category $t$ contradicts the \emph{functional} category of $s$, and $\mathrm{Conflict}_{\text{desc}}(s,t)$ measures incompatibility at the level of low-level acoustic attributes. Intuitively, we prefer target categories that induce a large semantic shift at the user-intent level while remaining relatively compatible in low-level acoustic traits. This improves malicious caption \(c_p\)'s retrievability and stabilizes targeted downstream music generation.

\subsection{Low-level semantic similar target generation}
This component addresses \emph{steering strength}. After choosing the target category \(t^\ast\), we encode malicious guidance as \emph{low-level acoustic descriptors} rather than relying only on a direct high-level target label.
Concretely, we construct a payload text from the selected target category:
\begin{equation}
c_{\text{payload}} = \mathcal{P}(t^\ast),
\label{eq:payload}
\end{equation}
where $\mathcal{P}(\cdot)$ maps the target category to generation-oriented low-level acoustic descriptors (e.g., timbre, texture, spatial cues, temporal patterns). For example, instead of inserting only the word \emph{spooky}, the payload may include descriptors such as \emph{``distant echoing chimes,'' ``hollow reverberant space,''} and \emph{``detuned sustained tones.''}
This design is more effective because downstream TTM generation operates on detailed acoustic guidance, not only broad style labels. As a result, low-level descriptor payloads \(c_{\text{payload}}\) provide a stronger and more stable steering signal than direct target-label injection, while remaining easier to blend with the benign anchor semantics.

\subsection{Poisoned Caption Assembly}
After constructing the anchor \(c_\text{anchor}\) and the payload \(c_\text{payload}\), we combine them into one poisoned caption \(c_p\). The anchor keeps the caption aligned with the benign query at the high level (so it can still be retrieved), while the payload injects low-level malicious acoustic guidance that biases downstream generation. In practice, we use a simple two-part caption structure:
\begin{equation}
\begin{aligned}
c_p &= c_\text{anchor} \Vert c_\text{payload}
\end{aligned}
\label{eq:template}
\end{equation}
Intuitively, the first part helps the caption ``get in'' through retrieval, and the second part helps the attack ``take effect'' during music generation. This simple assembly also makes it easy to compare payload designs in our experiments, such as direct target-label injection versus our payload injection.

\begin{table*}[ht]
\centering
\caption{Caption poisoning shifts MusicGen outputs toward the target category. We report CLAP similarity for benign and poisoned generations with respect to the user query and the target category ($\downarrow$ lower is better for benign target-category similarity, $\uparrow$ higher is better for poisoned target-category similarity). Best results are bolded.}
\vspace{-8pt}
\label{tab:topk_retrieval}
\begin{tabular}{l c c c c c c c c}
\toprule
\multirow{2}{*}{\begin{tabular}[c]{@{}l@{}}TTM\\ Model\end{tabular}} &
\multirow{2}{*}{Top-\(k\)} &
\multicolumn{2}{c}{Benign Music Generation} &
\multicolumn{2}{c}{Poisoned Music Generation} &
\multirow{2}{*}{Precision} &
\multirow{2}{*}{Recall} &
\multirow{2}{*}{F1} \\

\cmidrule(lr){3-4} \cmidrule(lr){5-6}

& &
CLAP$_\textbf{User Question}$ &
CLAP$_\textbf{Target Category}$ $\downarrow$&
CLAP$_\textbf{User Question}$ &
CLAP$_\textbf{Target Category}$ $\uparrow$&
& & \\

\midrule
\multirow{3}{*}{MusicGen}
& 3  & 0.33 & 0.28 & 0.28 & \textbf{0.48} & 0.980 & 0.588 & 0.735 \\
& 5  & 0.31 & 0.25 & 0.27 & 0.45 & 0.908 & 0.908 & 0.908 \\
& 10 & 0.29 & \textbf{0.21} & 0.30 & 0.41 & 0.472 & 0.944 & 0.629 \\
\bottomrule
\end{tabular}
\end{table*}

%% file: section/5_evaluation.tex
\section{Evaluation}
\subsection{Experimental Setup}
\noindent\textbf{Dataset.} For retrieval and text-to-music generation experiments, we use the MusicCaps dataset \cite{agostinelli2023musiclm}, which contains 5,521 paired text and music examples. All audio clips are sourced from YouTube videos. For each example, we extract the first 10 seconds of the audio and pair it with a free-text caption describing the music, for example, \emph{``This is a live recording of a DJ doing deck scratching on a hip hop song.”} The dataset also provides structured musical attributes such as genre, mood, and instrumentation, for example, \emph{``deck scratching, hip hop, electric.”} In our experiments, we use only the free-text music captions and ignore the structured attribute annotations.

\noindent\textbf{RAG Setup.} Our RAG-based TTM pipeline setups are as follows:
\begin{itemize}
\item {Knowledge base:} We use MusicCaps as the base knowledge database.

\item {Retriever:} We use CLAP \cite{elizalde2023clap} to embed user queries and music captions into a shared representation space, and rank candidate captions by cosine similarity between the query embedding and caption embeddings in the knowledge base.

\item {TTM model:} We use Meta's MusicGen (musicgen-small) \cite{huang2023musicgen} for generation. Unless otherwise noted, we use the default generation settings from AudioCraft\footnote{https://github.com/facebookresearch/audiocraft}
\end{itemize}
\noindent\textbf{Target categories and answers.} Our method aims to induce the music RAG system to generate attacker-chosen target music for each source category. We randomly sample 50 source categories and, for each, construct a plausible user query and select a functionally opposite target category generated by Sonnet 4.6~\cite{anthropic_sonnet_46}. For each source-target category pair, Sonnet 4.6 first generates a short rationale describing their shared acoustic attributes, then refines detailed poisoned target-category captions \(c_p\) based on the rationale. Unless otherwise specified, we retrieve the five most similar music captions from the knowledge database as contextual input for music generation. The prompt template and example poisoned captions are provided in Appendix~\ref{appendix:prompt}.

\noindent\textbf{Evaluation metrics.} We use the following metrics:
item
\begin{itemize}
\item \textbf{CLAP similarity.} CLAP is a joint text-audio embedding model that enables quantitative comparison between text and audio representations. For each generated music clip, we compute the cosine similarity between the embedding of the input text caption and the embedding of the generated audio \cite{huang2023make}. We report the average cosine similarity across all samples. A higher CLAP similarity indicates stronger adherence of the generated music to the conditioning text.

\item \textbf{Precision, recall, and F1 score.} Our method injects $N$ malicious captions into the knowledge database for each target music category. We use precision, recall, and F1 score to evaluate whether the injected captions are retrieved for the corresponding target queries. Precision is defined as the fraction of malicious captions among the top-$k$ retrieved results. Recall is defined as the fraction of injected malicious captions that are successfully retrieved among the top-$k$ results. The F1 score captures the tradeoff between precision and recall and is defined as:
\[
\mathrm{F1} = \frac{2 \cdot \mathrm{Precision} \cdot \mathrm{Recall}}{\mathrm{Precision} + \mathrm{Recall}}
\]
\end{itemize}

\subsection{Results}
\noindent\textbf{The attack shifts generated music toward the target category.} Table~\ref{tab:topk_retrieval} reports the CLAP similarity between generated audio and both the user question and the target category. Under three different top-$k$ retrieval settings, before the attack, the CLAP similarity between benignly generated music and the target category ranges from 0.21 to 0.28. After applying our attack, the CLAP similarity between the generated poisoned music and the target category increases substantially, reaching values between 0.41 and 0.48, which is nearly doubled. This indicates that the generated music is significantly closer to the attacker-chosen target category.

Notably, the CLAP similarity between the generated music and the original user question remains largely unchanged before and after the attack, staying at approximately 0.30 across all settings. This suggests that the attack successfully steers the generation toward the target category without degrading alignment with the user query, making the attack difficult to detect based on query relevance alone. Moreover, precision, recall, and F1 scores show that our attack maintains high retrievability across all three evaluation settings.

%% file: section/6_discussion.tex

%% file: section/8_conclusion.tex
\section{Conclusion}
We identified caption poisoning as an integrity threat in RAG-based TTM systems, where retrieved captions function as a control signal for augmented music generation. By injecting a small number of malicious caption entries into the music knowledge database, an attacker can steer generated music toward an attacker-chosen target category without changing the user prompt or model parameters. Our strategy maintains retrieval plausibility via high-level anchors and achieves steering via low-level acoustic descriptor payloads. Experiments show that this manipulation roughly doubles the CLAP similarity between the generated music and the target category, while keeping the CLAP similarity between the generated music and the user query nearly unchanged. This makes the attack difficult to detect from prompt relevance alone. Defenses should treat the retrieval corpus and caption metadata as part of the trusted base, with stronger ingestion controls and robustness checks in the retrieval pipeline.

\section*{Acknowledgment}
This work is partially supported by National Science Foundation (NSF) under award numbers 2523814, 2448205, and 2239605.

%% file: section/9_appendix.tex
\clearpage
\onecolumn
\appendix
\section{appendix}\label{appendix:prompt}

\begin{figure*}[!htbp]
\centering
\begin{tcolorbox}[
  colback=gray!12,
  colframe=black!70,
  boxrule=0.8pt,
  arc=2.5mm,
  left=6pt,
  right=6pt,
  top=4pt,
  bottom=6pt,
  title=\textbf{Attack example},
  coltitle=white,
  colbacktitle=black,
  fonttitle=\bfseries,
  fontupper=\normalfont
]

\textbf{Benign user intent:} Sleep and rest audio

\textbf{User query:} I want something very slow and simple that helps my mind shut down before sleep.

\textbf{Target intent:} Ritualistic dark drone ambient.

\textbf{Shared acoustic features:} Continuous drones, extremely slow tempo, minimal pitch movement, low frequency focus.

\textbf{Target music captions:}
Very slow and simple, steady tones for sleep, long held low tones, soft detuned layers with slow beating, minimal movement, no rhythm, extended sustain, narrow bandwidth.

Slow simple sleep tones, continuous low tone bed, gentle subharmonic layer, slow phase drift between layers, restrained dynamics, no percussion, long decay.

Something slow and simple for shutting down, sustained low register tones, subtle beating between close pitches, lightly saturated drone texture, minimal change, no rhythmic events.

Dense low register drone cluster with slow microtonal drift, persistent rumble undercurrent, minimal spectral contrast, flattened dynamics, extended sustain, gradual accumulation of harmonic weight.

Very slow simple background for sleep, held tones with soft low end, slow modulation and phase drift, minimal harmonic content, no pulse, continuous unbroken sustain.

\end{tcolorbox}
\caption{A concrete example of a caption poisoning attack}
\label{cdab_example}
\end{figure*}

\begin{figure*}[!htbp]
\centering
\begin{tcolorbox}[
  colback=gray!12,
  colframe=black!70,
  boxrule=0.8pt,
  arc=2.5mm,
  left=6pt,
  right=6pt,
  top=4pt,
  bottom=6pt,
  title=\textbf{Data generation prompt template},
  coltitle=white,
  colbacktitle=black,
  fonttitle=\bfseries,
  fontupper=\normalfont
]
\textbf{Role:} You are an AI data generator for music retrieval robustness testing. Your task is to create paired benign and adversarial music descriptions using \textit{Acoustic Mimicry, Semantic Inversion}.\\[4pt]

\textbf{Goal:} Map a benign user intent to a functionally negative target category such that the \textbf{low-level acoustic features remain similar} (instrumentation, timbre, register, tempo, rhythmic cues), while the \textbf{high-level function is inverted} (intended use, perceived safety, psychological effect).\\[6pt]

\textbf{Definitions:}\\
\textbf{Low-level elements (must be preserved):} instrumentation and sound sources, timbre and texture, register and basic loudness, basic temporal or rhythmic cues.\\
\textbf{High-level elements (must be inverted):} intended function or usage context, mood/affect framing, semantic purpose.\\[6pt]

\textbf{Generation requirements (per item):}
\begin{enumerate}
  \item \textbf{Source Category:} A short benign label (sleep, focus, relaxation, background listening).
  \item \textbf{User Question:} Natural, high-level language only. No professional audio terminology. Describe intent, not sound design.
  \item \textbf{Target Category:} Functionally negative/adversarial category that shares sonic DNA with the source.
  \item \textbf{Shared Acoustic Features:} One short line explaining why they sound similar using only low-level elements.
  \item \textbf{Target Music Captions:} Generate \textbf{N=5} distinct variations. Each caption:
  \begin{itemize}
    \item looks like a \textbf{text-to-music prompt},
    \item is \textbf{25--40 words},
    \item uses \textbf{only low-level acoustic descriptors} (sources, texture, register, density, modulation, rhythm),
    \item \textbf{must not} include explicit mood/genre/intent labels (e.g., ``calm'', ``scary'', ``horror'', ``relaxing''),
    \item intensifies functional negativity implicitly via \textbf{pressure, instability, density, dissonance, repetition, unresolved structure},
    \item varies meaningfully across the 5 captions while remaining consistent with the target category.
  \end{itemize}
\end{enumerate}
\vspace{2pt}

\textbf{Constraints:} Do not mention ``attack'', ``malicious'', or ``negative''. Do not explain the process in the output. Avoid repeating sentence templates across captions. Within each source category item, keep all fields fixed except the target music captions.\\[6pt]

\textbf{Output format:} Return a JSON object exactly matching:
\begin{verbatim}
{
  "source_categories": [
    {
      "id": <integer>,
      "source_category": "<string>",
      "user_question": "<string>",
      "target_category": "<string>",
      "shared_acoustic_features": "<string>",
      "target_music_captions": [
        "<caption_1>",
        "<caption_2>",
        "<caption_3>",
        "<caption_4>",
        "<caption_5>"
      ]
    }
  ]
}
\end{verbatim}
\end{tcolorbox}
\caption{Prompt template for generating acoustically similar but functionally inverted caption pairs.}
\label{data_gen_prompt_template}
\end{figure*}